\documentclass[10pt,twocolumn]{article}
\usepackage[a4paper,margin=1.8cm,columnsep=0.6cm]{geometry}
\usepackage{pdfpages}
\usepackage{graphicx}
\usepackage{amsmath,amssymb}
\usepackage{booktabs}
\usepackage{siunitx}
\usepackage[numbers,sort&compress,super]{natbib}
\usepackage{xcolor}
\usepackage{xspace}
\usepackage[hidelinks]{hyperref}
\usepackage{caption}
\usepackage{tabularx}

\newcommand{\todo}[1]{\textcolor{red}{[TODO: #1]}}
\newcommand{\figfile}[2]{\IfFileExists{#2}{\includegraphics[width=#1]{#2}}{\fbox{\parbox{0.9\linewidth}{\centering\todo{figure \texttt{\detokenize{#2}} pending}}}}}
\newcommand{\SphereRmsDb}{\SI{0.11}{\decibel}\xspace}
\newcommand{\SphereMaxDb}{\SI{0.53}{\decibel}\xspace}
\newcommand{\SphereRmsDeg}{\SI{0.5}{\degree}\xspace}
\newcommand{\IrregCbie}{\SI{14.4}{\decibel}\xspace}
\newcommand{\IrregBm}{\SI{0.03}{\decibel}\xspace}
\newcommand{\NumcalcMaxRms}{\SI{0.13}{\decibel}\xspace}
\newcommand{\RefPdiscL}{0.76\xspace}
\newcommand{\RefPdiscR}{0.82\xspace}
\newcommand{\RefPdiscE}{0.84\xspace}
\newcommand{\RefPdiscSmooth}{0.927\xspace}
\newcommand{\RefMmicrL}{0.92\xspace}
\newcommand{\RefMmicrR}{0.93\xspace}
\newcommand{\RefMmicrE}{0.96\xspace}
\newcommand{\RefMmicrSmooth}{0.998\xspace}

\newcommand{\InterpRmsMax}{\SI{0.77}{\decibel}\xspace}

\newcommand{\PistonDiTwenty}{1.5--1.9\xspace}

\newcommand{\EnsEarCorr}{0.75--1.00\xspace}
\newcommand{\EnsEmCorr}{0.15--0.83\xspace}
\newcommand{\ItdCorr}{0.65--0.97\xspace}

\newcommand{\EmDelayCorr}{0.95--1.00\xspace}
\newcommand{\EmDelayCorrRform}{0.13\xspace}

\newcommand{\FarCorr}{0.992\xspace}
\newcommand{\FarErr}{0.33\xspace}
\newcommand{\VhCorrOurs}{0.70\xspace}
\newcommand{\VhCorrTheirs}{0.69\xspace}
\newcommand{\VhCorrSim}{0.997\xspace}

\newcommand{\HbHalfDiff}{\SI{15}{\degree}\xspace}

\newcommand{\HbRadius}{\SI{3.5}{\milli\meter}\xspace}

\title{Full-sphere emission and reception directivities of seven echolocating bat species}
\author{Jan Steckel$^{1,2}$\\[0.4em]\small $^1$Cosys-Lab, University of Antwerp, Antwerp, Belgium\\ \small $^2$Flanders Make, Lommel, Belgium\\ \small Corresponding author: Jan Steckel (jan.steckel@uantwerpen.be)}
\date{}

\begin{document}
\twocolumn[{\maketitle\vspace{-1.5em}\begin{center}\begin{minipage}{0.86\textwidth}\small\noindent\textbf{Abstract.} Echolocating bats shape their calls and the returning echoes with their noseleaf or mouth, head and pinnae. Understanding these spectrospatial directivity patterns is central to bat echolocation. We describe a dataset of simulated far-field emission and reception directivities of seven echolocating bat species from three families (Phyllostomidae, Rhinolophidae and Vespertilionidae). For each species, the head of one preserved specimen was scanned with micro-computed tomography and converted into a closed surface mesh, on which the responses of both ears and of the emitter were computed with a boundary element method or, for the two mouth-emitting species, with an analytical piston model. The complex-valued responses cover the full sphere at \SI{2.5}{\degree} resolution from 20 to \SI{160}{\kilo\hertz} in \SI{0.5}{\kilo\hertz} steps. They come with the meshes, the landmarks of ears, nostrils and mouth, alternative source positions and skeletons for posing the pinnae and noseleaves, all in one frame, and with Python and MATLAB code. The technical validation covers solver verification, mesh convergence, comparisons with earlier simulations and measurements, and sensitivity analyses.\end{minipage}\end{center}\vspace{1.5em}}]

\section*{Background \& Summary}
Echolocating bats perceive their surroundings through the echoes of their own calls, and the acoustics of their heads are an integral part of this sensing system. Indeed, the emission beam is shaped by the mouth or by the noseleaf \citep{zhuangNoseleafFurrowsHorseshoe2006, vanderelstWhatNoseleavesFM2010, vanderelst2013}, and the pinnae and the head impose a strongly direction- and frequency-dependent filter on the received echoes \citep{wotton1995, aytekin2004, firzlaff2003, demeySimulatedHeadRelated2008}. For the receiver, this filter is the Head-Related Transfer Function (HRTF) known from human hearing. The product of the emission and reception directivities is called the Echolocation-Related Transfer Function (ERTF) \citep{demeySimulatedHeadRelated2008, vanderelstWhatNoseleavesFM2010}, which describes the spatial sensitivity of the complete sonar system. Directivities of this kind are the starting point for models of echo-based localization and of the information available to the bat \citep{reijniers2010, vanderelst2011}, for simulations of biosonar in complex environments \citep{steckel2013}, and for comparative studies of how morphology shapes directivity \citep{obrist1993, ma2011}.

Measuring these directivities is, however, a tedious process. The emission and the hearing directivity of a bat have to be sampled over hundreds to thousands of directions and over a frequency band that easily spans 20 to \SI{150}{\kilo\hertz}, either on live animals, whose calls, ear postures and head positions vary \citep{ghose2003, linnenschmidt2016, hafele2026}, or on dead specimens, whose tissue deforms \citep{aytekin2004, vesterholm2025}. Moreover, most of the published measurements are available only as figures, not as data \citep{obrist1993, wotton1995, firzlaff2003, firzlaff2004, aytekin2004}. Numerical simulation of the sound field around a three-dimensional head model offers an alternative, and the Boundary Element Method (BEM) has been used for this purpose in bats for almost twenty years \citep{zhuangNoseleafFurrowsHorseshoe2006, demeySimulatedHeadRelated2008, muller2008, vanderelstWhatNoseleavesFM2010, vanderelst2011, ma2011, vanderelst2013}. These studies, however, mostly reported magnitudes, often on the frontal hemisphere only and for one species at a time, and their transfer functions were not released as datasets. To our knowledge, the public data closest to the dataset described here are the measured and simulated HRTFs of \textit{Myotis daubentonii} and \textit{Myotis nattereri}, with the corresponding head meshes, of Vesterholm et al. \citep{vesterholm2025, vesterholm2025data} (one ear, up to \SI{130}{\kilo\hertz}), and the simulated HRTFs of live \textit{M. daubentonii} and the in-flight emission beam of that species of H\"afele et al. \citep{hafele2026, hafele2026data}.

In this Data Descriptor, we describe a dataset of simulated emission and reception directivities of seven species: \textit{Phyllostomus discolor}, \textit{Micronycteris microtis}, \textit{Macrophyllum macrophyllum}, \textit{Rhinolophus rouxii}, \textit{Rhinolophus formosae}, \textit{Eptesicus fuscus} and \textit{Myotis daubentonii} (Table~\ref{tab:species}). For each species, it contains (1) a closed, simulation-ready head mesh derived from a micro-computed tomography (micro-CT) scan, (2) the anatomical landmarks used as sources: the ear canal entrances, and the nostrils or the mouth, (3) the complex-valued far-field transfer functions of both ears and of the emitter on the full sphere from 20 to \SI{160}{\kilo\hertz}, (4) the transfer functions of up to four alternative positions of every ear and of the nostril pair, and (5) a skeleton for the pinnae and the noseleaf, with which the meshes can be posed and simulated again. Meshes, landmarks and directivities share one coordinate frame, so that they are spatially aligned without any further registration. The emission patterns of the five nasal emitters are BEM results, whereas the emitters of the two mouth-emitting vespertilionids are analytical piston models calibrated to published beam widths, with alternative radii, as their beam depends on the task of the bat, and we keep this distinction throughout the data and this paper. All responses were newly computed for this dataset with a GPU implementation of the BEM solver, which we verified against the open-source solver NumCalc of Mesh2HRTF \citep{ziegelwangerMesh2HRTFOpensourceSoftware2015, kreuzerFastMultipoleBoundary2009, brinkmann2023}. The meshes of \textit{P. discolor} and \textit{M. microtis} were used before, at lower resolution and with another solver, for the simulations of De Mey et al. \citep{demeySimulatedHeadRelated2008} and Vanderelst et al. \citep{vanderelstWhatNoseleavesFM2010}. The meshes of \textit{R. rouxii} \citep{vanderelst2011} and \textit{R. formosae} \citep{vanderelst2013} were also used in earlier studies. Those of \textit{M. macrophyllum}, \textit{E. fuscus} and \textit{M. daubentonii} are published here for the first time. The released meshes are new remeshes of the raw CT surfaces, with modern mesh-cleanup techniques applied.

As every species is represented by a single specimen, the dataset is a collection of specimen-based reference models, not of species averages, and it does not capture variation within a species. Furthermore, the responses are far-field directivities of acoustically rigid heads, not measurements and not full solutions of the near field. With these limits in mind, we expect the data to be useful for the simulation of bat biosonar and of echo acoustics, for localization and head-movement models, for comparative studies of morphology and directivity, as training data for acoustic learning, and for the design of biomimetic sonar. The code that computes the responses, together with a small simulator that turns them into binaural echo signals, is released alongside the data.

\section*{Methods}
Fig.~\ref{fig:frames} shows the seven head models with their landmarks and coordinate frames. In what follows, we describe the specimens and meshes, the coordinate frame, the acoustic model and the solver, the source models and their placement, and the evaluation of the far field. Throughout the derivations, we use the $e^{-j\omega t}$ time convention of the solver. The dataset itself uses the $e^{+j\omega t}$ convention, and the conversion (a complex conjugation) is part of the referencing in equation~\ref{eq:ref}.

\begin{table*}[!tbp]
\centering
\caption{Species, Specimens and Meshes. Head span: largest extent of the mesh. Faces and mean edge length of the clean meshes. $f_6$: frequency with six mean-size elements per wavelength. Deviation: median distance from points on the clean mesh to the original CT surface. Band: computed frequency range (\SI{0.5}{\kilo\hertz} steps). Emitter: nasal (two vibrating nostril patches, BEM) or oral (analytical piston of radius $a$, calibrated to the open-field search beam, with five alternative radii from 2.5 to \SI{6.5}{\milli\meter} in the dataset). \textit{M. microtis} is spelled \textit{microtus} in the original mesh files and in the dataset file names. One preserved specimen per species.}
\label{tab:species}
\resizebox{\linewidth}{!}{\begin{tabular}{llrrrrrll}
\toprule
Species & Family & Span (mm) & Faces & Edge (mm) & $f_6$ (kHz) & Deviation (mm) & Band (kHz) & Emitter \\
\midrule
\textit{Phyllostomus discolor} & Phyllostomidae & 39 & 53{,}984 & 0.392 & 146 & 0.052 & 20--145 & nasal \\
\textit{Micronycteris microtis} & Phyllostomidae & 27 & 51{,}980 & 0.270 & 211 & 0.024 & 20--160 & nasal \\
\textit{Macrophyllum macrophyllum} & Phyllostomidae & 23 & 46{,}306 & 0.277 & 207 & 0.049 & 20--160 & nasal \\
\textit{Rhinolophus rouxii} & Rhinolophidae & 25 & 53{,}672 & 0.279 & 205 & 0.047 & 20--160 & nasal \\
\textit{Rhinolophus formosae} & Rhinolophidae & 40 & 57{,}746 & 0.424 & 135 & 0.061 & 20--135 & nasal \\
\textit{Eptesicus fuscus} & Vespertilionidae & 30 & 52{,}474 & 0.299 & 191 & 0.053 & 20--160 & oral{,} $a$ = 6.59 mm \\
\textit{Myotis daubentonii} & Vespertilionidae & 22 & 35{,}536 & 0.276 & 207 & 0.054 & 20--160 & oral{,} $a$ = 6.28 mm \\
\bottomrule
\end{tabular}
}
\end{table*}

\subsection*{Specimens and head meshes}
The head meshes are legacy meshes from the archive of the University of Antwerp, made from micro-CT scans of one preserved specimen per species with the procedure of De Mey et al. \citep{demeySimulatedHeadRelated2008}. These scans are a valuable resource. Bat specimens suitable for micro-CT are scarce, every new scan requires a new animal specimen, and the scanning and segmentation of a head take considerable effort. By releasing the meshes together with their simulated directivities and the tools to pose and simulate them again, we want to make the most of these animal specimens. In this way, the specimens can serve many future studies without new animal material, and this resource is preserved for future research. For \textit{P. discolor} and \textit{M. microtis}, the heads were scanned with a SkyScan 1076 micro-CT scanner at a resolution of \SI{35}{\micro\meter}, the reconstructed images were downsampled to \SI{70}{\micro\meter}, and the tissue was segmented from the background with Amira, automatically where possible and with manual corrections \citep{vanderelstWhatNoseleavesFM2010}. For \textit{R. rouxii}, a scan resolution of \SI{70}{\micro\meter} is reported \citep{vanderelst2011}. The segmented surfaces contain the head, including the pinnae with tragus and, where present, the noseleaf, truncated behind the skull, not containing body or wings such as in previous work \citep{demeySimulatedHeadRelated2008}. The pinnae are in the posture in which the specimens were preserved, and the mouths of all specimens are closed (Supplementary Fig.~S1).

As the BEM requires a closed, consistently oriented surface, we remeshed every CT mesh with isotropic explicit remeshing in PyMeshLab \citep{pymeshlab}, at a target edge length of \SI{0.27}{\milli\meter}, or longer where needed to stay within 58,000 triangles, and repaired the remaining defects locally until each mesh was closed, two-manifold, a single connected body and free of self-intersections (Supplementary Note 1). The clean surfaces deviate from the CT surfaces by 0.02 to \SI{0.06}{\milli\meter} (median). Table~\ref{tab:species} lists the meshes. For \textit{P. discolor} and \textit{R. formosae}, whose heads are larger, the face budget forced longer edges, and we limited their bands to 145 and \SI{135}{\kilo\hertz}, where the meshes still have six elements per wavelength.

\subsection*{Coordinate frame and landmarks}
\begin{figure*}[!htbp]
\centering
\figfile{\linewidth}{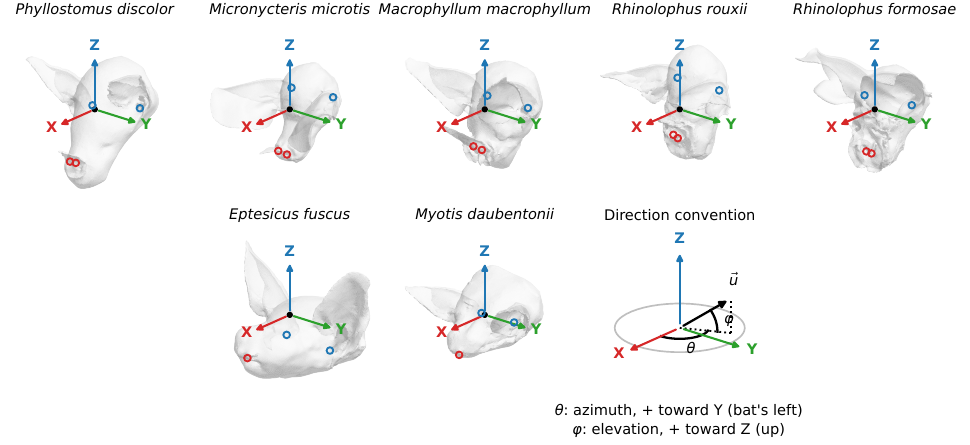}
\caption{Head Meshes Posed in the Coordinate Frame of the Transfer Functions. Each panel shows a mesh exactly as it was simulated, seen from front-left-above, with the origin (black dot, the centre of the evaluation sphere) and the axes $X$ (red, azimuth \SI{0}{\degree}), $Y$ (green, azimuth \SI{+90}{\degree}, the bat's left) and $Z$ (blue, elevation \SI{+90}{\degree}). Blue circles mark the ear receiver patches, red circles the nostril patches or the centre of the mouth piston. Panels are labelled by species. The sketch shows the azimuth $\theta$ and elevation $\varphi$ of a direction $\vec{u}$. The pitch of the frame differs between species, as it follows the orientation of the original scans, except for \textit{P. discolor}, whose head is shown in the pose of its earlier simulation \citep{demeySimulatedHeadRelated2008}.}
\label{fig:frames}
\end{figure*}

All data are given in the frame of the original CT meshes (for \textit{P. discolor} in a rotated pose, see below), with the axes renamed: $X$ points forward, $Y$ to the bat's left and $Z$ up, in metres. A direction $\psi$ is given by its azimuth $\theta$ (positive to the bat's left) and elevation $\varphi$ (positive up), with unit vector:
\begin{equation}
\vec{u}(\psi) = \left[\cos\varphi \cdot \cos\theta,\ \cos\varphi \cdot \sin\theta,\ \sin\varphi\right]^T.
\end{equation}
This frame is the frame of the original simulations: when we fitted the head orientation of \textit{M. microtis} to the reference ERTFs of Vanderelst et al. \citep{vanderelstWhatNoseleavesFM2010}, using only the acoustic data, the fitted frame coincided with the frame of the CT file to within \SIrange{2}{5}{\degree}. For the nasal emitters, this frame orients the plane of the nostril openings approximately vertically \citep{vanderelstWhatNoseleavesFM2010}. For the other species, it follows the orientation of the scan, so that the pitch of the head differs between species (Fig.~\ref{fig:frames}). We therefore did not rotate the data into a common anatomical frame, which would require choices (for instance, of a skull-based horizontal plane) that are better left to the user. As the data cover the full sphere, they can be rotated into any other frame without loss. The one exception is \textit{P. discolor}, whose head we rotated by \SI{33}{\degree} from its CT frame into the pose of its earlier simulation \citep{demeySimulatedHeadRelated2008}, in which the ERTFs of this species are usually shown. Its mesh, landmarks and transfer functions are all given in this pose. For every species, the mesh is given in exactly the coordinates of its transfer functions (Fig.~\ref{fig:frames}), and the file names keep the suffix \texttt{\_ct} also for \textit{P. discolor}, whose rotation from the CT scan is recorded in \texttt{landmarks.json}.

The landmarks are the centres of the vibrating patches described below: one per ear canal entrance, one per nostril for the nasal emitters, and the mouth, on the midline at the lip seam, for the oral emitters. Their placement is described in the section on source placement.

\subsection*{Acoustic model and solver}
We model the head as acoustically rigid, with the sound sources as small vibrating patches on its surface, in air with a speed of sound of \SI{343}{\meter\per\second} and a density of \SI{1.1839}{\kilo\gram\per\meter\cubed}. The sound field is computed with the BEM in the Burton-Miller formulation \citep{burtonApplicationIntegralEquation1971}, which, in contrast to the conventional boundary integral equation, has a unique solution at all frequencies, discretized with constant triangular elements and collocation at the element centroids, as in NumCalc of Mesh2HRTF \citep{kreuzerFastMultipoleBoundary2009, ziegelwangerMesh2HRTFOpensourceSoftware2015}. Our implementation assembles the dense system on a GPU and solves it per frequency by an LU decomposition, after which every additional source (both ears, the emitter and all alternative positions) costs little. Supplementary Note 2 gives the formulation, the quadrature and the computational details. The solver is part of the released code \citep{thiscode}, and its version is stored in the metadata of every file.

\subsection*{Sources: ears and emitters}
The transfer function of an ear is computed by acoustic reciprocity: a small patch at the ear canal entrance (radius \SI{0.4}{\milli\meter}) vibrates with uniform normal velocity, and the pressure it radiates to a far-field point, normalized as described below, equals the patch-averaged pressure that a source with the same volume velocity at that point produces at the ear. This is the approach of Mesh2HRTF \citep{ziegelwangerMesh2HRTFOpensourceSoftware2015}, and it is the patch equivalent of the four averaged point receivers of the earlier bat simulations \citep{demeySimulatedHeadRelated2008, vanderelstWhatNoseleavesFM2010}. The ear canal itself and the middle ear are not modelled. On the coarser meshes of \textit{P. discolor} and \textit{R. formosae}, the source patches are resolved by only 4 to 11 elements (14 to 26 for the other species, Supplementary Note 3). The nasal emitters (\textit{P. discolor}, \textit{M. microtis}, \textit{M. macrophyllum} and both \textit{Rhinolophus} species) are modelled as two vibrating patches (radius \SI{0.3}{\milli\meter}), one per nostril (smaller than the real nostril openings, to keep the source position well defined), with equal normal velocity and phase, so that the interference between the nostrils, which is central to the beam forming of the noseleaf \citep{zhuangNoseleafFurrowsHorseshoe2006, vanderelstWhatNoseleavesFM2010}, is included.

\textit{E. fuscus} and \textit{M. daubentonii} emit their calls through the open mouth, whose gape acts as the radiating aperture and which the bats adjust during echolocation \citep{jakobsenVespertilionidBatsControl2010}. The preserved specimens, and thus the meshes, have closed mouths. A vibrating patch on the closed snout therefore has no aperture to shape the beam: patches of up to \SI{4}{\milli\meter} radius gave half-amplitude angles of 33 to \SI{52}{\degree}, against 14 to \SI{30}{\degree} measured in the field (Supplementary Note 3). Simulating the open mouth would require a mesh with a modelled gape, which the specimens do not provide. We therefore model the emission of these two species with an analytical circular piston \citep{surlykkeEcholocatingBatsEmit2009, jakobsenConvergentAcousticField2013}. The piston approximates the open mouth and does not include the head. Referenced as in equation~\ref{eq:ref} and in the $e^{+j\omega t}$ convention of the dataset, it reads as follows:
\begin{equation}
H_E(\psi, f) = 2 \cdot \frac{2 J_1(k a \sin\gamma)}{k a \sin\gamma} \cdot \frac{1 + \cos\gamma}{2} \cdot e^{j k\, \vec{u}(\psi) \cdot \vec{p}_m},
\label{eq:piston}
\end{equation}
where $k$ is the wavenumber, $\gamma$ the angle between $\vec{u}(\psi)$ and the piston axis (the mean outward surface normal around the mouth), $J_1$ the Bessel function of the first kind and first order, $a$ the piston radius, and $\vec{p}_m$ the mouth landmark. The piston contains no head, so that its rear hemisphere, and the two-way ERTF built from it, cannot be compared with those of the BEM emitters. We calibrated the radius to published field measurements of the beam width: \SI{6.28}{\milli\meter} for \textit{M. daubentonii} \citep{surlykkeEcholocatingBatsEmit2009} and \SI{6.59}{\milli\meter} for \textit{E. fuscus} \citep{hulgardBigBrownBats2016}. These calibrations describe search calls emitted in free flight in the open field, which form the narrowest beam these bats emit. Mouth-emitting bats widen their beam in the laboratory and in confined spaces \citep{surlykkeEcholocatingBatsEmit2009, jakobsenVespertilionidBatsControl2010, hartley1989}, so a single radius cannot represent all behaviours. We therefore treat the radius as a parameter of the emitter model. The dataset contains the piston for five radii from 2.5 to \SI{6.5}{\milli\meter} in steps of \SI{1}{\milli\meter}, which span the published beams of these species from the laboratory (3.1 to \SI{4.7}{\milli\meter} \citep{surlykkeEcholocatingBatsEmit2009, hartley1989}) to the open field, and the released code computes the piston for any radius. Supplementary Note 3 gives the details of the source models and of the calibration.

\subsection*{Source placement}
Where exactly the receiver sits at the ear canal entrance, and where the nostrils are, is a modelling choice that the meshes do not fix, and our comparisons with the earlier simulations showed that it matters (see Technical Validation). To quantify this, we computed dense sets of candidate sources for the two species with earlier reference simulations, \textit{P. discolor} \citep{demeySimulatedHeadRelated2008} and \textit{M. microtis} \citep{vanderelstWhatNoseleavesFM2010}, and compared each with the reference (Supplementary Fig.~S3). For these two species, the landmarks are the best-matching receiver positions and nostril pair. To place the sources of the other five species, we derived a rule from these two: the best-matching positions are those that radiate most strongly forward (front/back ratio) and are most occluded by the surrounding anatomy, which is where pinna and noseleaf focus sound forward.

Each receiver is placed at the maximum of the standardized sum of the front/back ratio and the occlusion, and the nostrils at the mirror-symmetric pair 1.5 to \SI{2.5}{\milli\meter} apart with the highest front/back ratio. On the two calibration species, the rule reaches pattern correlations of 0.77 to 0.86 for the ears (means over the ten placement frequencies), against 0.78 to 0.94 for the best possible position and 0.54 to 0.70 for the median candidate, and 0.75 to 0.81 when the rule is chosen without the held-out species. Tested against the measured HRTF of another \textit{M. daubentonii} specimen, the rule places the receiver within 0.01 in pattern correlation of the best of 211 candidates (Technical Validation). For \textit{R. formosae}, we deviated from the rule for the nostrils. Its first nostril pair lies on the internarial region at the base of the sella and produces a beam with several lobes of nearly equal level at 36 to \SI{43}{\kilo\hertz}, the range of its calls, whereas the earlier simulation of the noseleaf of this species showed a single main lobe (Fig.~3A of Vanderelst et al. \citep{vanderelst2013}). The second pair of the rule lies in the nostril openings within the horseshoe and produces a single main lobe, with the next lobe 4.6 to \SI{7.3}{\decibel} lower. We therefore use this pair as the nominal emitter, and the first one as the first alternative. As the rule was derived from only four ears and two emitters, every species also contains up to four alternative positions per ear and per nostril pair, 0.5 to \SI{5}{\milli\meter} from the nominal ones, which show how strongly a result depends on this choice. The mouth landmark of the oral emitters was placed by hand on the midline at the lip seam. Supplementary Note 4 and Supplementary Table~S1 give the complete procedure, its parameters and the results.

\subsection*{Posable head models}
To allow studies of how the posture of the pinnae and the noseleaf shapes the directivities, every mesh comes with a skeleton (see the Technical Validation for an example). The pinnae and the noseleaves were segmented automatically as the thin structures that disappear when the head volume is morphologically opened with a ball of \SI{1}{\milli\meter} radius, and they were assigned to the ears and the nose with the landmarks. Each pinna carries a chain of three bones from its base to its tip, and each noseleaf of the five nasal species a chain of two, all attached to the fixed head. Every bone rotates about four axes: flex (backward or forward), abduct (outward or inward), yaw (about the vertical axis of the head, turning the opening of a pinna forward or backward) and twist (about the bone itself). The surface follows the bones through smooth skinning weights and dual-quaternion skinning. A posed mesh therefore has the same triangles as the rest mesh, so that the source patches and landmarks carry over. A validity check rejects poses that make the surface intersect itself, invert faces or stretch edges by more than a factor of two. The rest pose is the preserved posture of the specimen, for which the transfer functions of the dataset were computed. Supplementary Note 8 gives the construction and the valid angle range of every joint.

\subsection*{Far field, frequencies and referencing}
The field is evaluated on a sphere of radius $R = \SI{1}{\meter}$ around the origin of the frame, at azimuths $-180$ to \SI{177.5}{\degree} and elevations $-90$ to \SI{90}{\degree} in \SI{2.5}{\degree} steps (73 by 144 directions, the poles are computed once and repeated), from 20 to \SI{160}{\kilo\hertz} in \SI{0.5}{\kilo\hertz} steps (281 frequencies, 251 for \textit{P. discolor} and 231 for \textit{R. formosae}). At this distance, the magnitude patterns equal the far-field directivity to within pattern correlations of 0.99, while the phase still contains the curvature of the wavefront over the head (Supplementary Note 5), and we describe the data as far-field directivities evaluated at \SI{1}{\meter}. Every transfer function is referenced to the free-field pressure at the same point of a monopole at the origin with the same volume velocity $Q$ as the source:
\begin{equation}
H(\psi, f) = \left[\frac{p(R\,\vec{u}(\psi))}{p_{ff}(R)}\right]^*, \qquad p_{ff}(R) = -j\omega\rho Q \cdot \frac{e^{jkR}}{4\pi R},
\label{eq:ref}
\end{equation}
as in Mesh2HRTF \citep{ziegelwangerMesh2HRTFOpensourceSoftware2015}, where the complex conjugate converts the result into the $e^{+j\omega t}$ convention of the discrete Fourier transform. With this referencing, the spherical spreading and the propagation delay from the origin are removed, and the phase of $H$ is the phase relative to a wave arriving at (or leaving from) the origin. An inverse FFT over frequency therefore yields an impulse response with $t = 0$ at the origin, which, for directions on the side of the source, starts up to $|\vec{p}|/c$ (about \SI{60}{\micro\second}) before $t = 0$. The magnitude of $H$ is the pressure gain relative to the same source in free field. $|H| \approx 1$ at low frequencies for a source far from any obstacle, and $|H| = 2$ on the axis of a baffled piston.

\section*{Data Records}
The dataset is available at Zenodo \citep{thisdataset} (\url{https://doi.org/10.5281/zenodo.23729269}), under a Creative Commons Attribution 4.0 International (CC BY 4.0) licence. It is organized in one folder per species, named after the species as in the original mesh files (for example, \texttt{Micronycteris\_microtus} for \textit{M. microtis}), plus, at the top level, a \texttt{README.md} with the conventions, \texttt{species.json} (folder, species, family and emitter type of every species), \texttt{meshes.json} (mesh statistics and deviation from the CT surface), \texttt{checksums.txt} (SHA-256 checksum and size of every file), and \texttt{code/}, which contains the Python module to load and use the data (\texttt{batertf/dataset.py}, which needs only NumPy and h5py) and examples in Python and MATLAB. Table~\ref{tab:files} lists the files of a species folder and Table~\ref{tab:hdf5} the content of the HDF5 files. On Zenodo, every species folder is one zip archive of the same name, the loader and the examples are in \texttt{code.zip}, and the complete code that produced the dataset is in \texttt{bat\_ertf\_code-1.0.0.zip} (see Code Availability). \texttt{checksums.txt} lists the unpacked files. The total size of the dataset is \SI{3.3}{\giga\byte}.

\begin{table*}[!tbp]
\centering
\caption{Files per Species Folder.}
\label{tab:files}
\small
\renewcommand{\arraystretch}{1.15}
\begin{tabularx}{\linewidth}{>{\raggedright}p{0.25\linewidth} >{\raggedright\arraybackslash}X}
\toprule
File & Content \\
\midrule
\texttt{mesh.stl} & Clean head mesh (closed, two-manifold, one body), binary STL, metres, in the coordinates of \texttt{ertf\_ct.h5}. \\
\texttt{ertf\_ct.h5} & All transfer functions (nominal and alternative sources) with metadata (Table~\ref{tab:hdf5}). \\
\texttt{ears\_ct.sofa} & Nominal left and right ear, SOFA \texttt{SimpleFreeFieldHRTF} \citep{aes69_2022, majdakSpatiallyOrientedFormat2013}, 10,512 directions. \\
\texttt{emitter\_ct.sofa} & Nominal emitter, SOFA \texttt{FreeFieldDirectivityTF}. \\
\texttt{HRTF\_<species>.mat} & Nominal ears and emitter in the MATLAB format of the earlier simulations \citep{demeySimulatedHeadRelated2008, vanderelstWhatNoseleavesFM2010} (struct \texttt{bat\_hrtf}). \\
\texttt{landmarks.json} & Species, emitter type, landmarks, alternative source positions, piston parameters, frame rotations, mesh statistics and conventions. \\
\texttt{rig/} & Posable model: skeleton (\texttt{rig.json}), skinning weights (\texttt{weights.npz}) and rigged glTF model (\texttt{<species>.glb}). \\
\bottomrule
\end{tabularx}
\end{table*}

\begin{table*}[!tbp]
\centering
\caption{Content of the HDF5 Files (\texttt{ertf\_<frame>.h5}).}
\label{tab:hdf5}
\small
\begin{tabularx}{\linewidth}{lX}
\toprule
Path & Content \\
\midrule
\texttt{freqs} & Frequencies (Hz), \SI{0.5}{\kilo\hertz} steps. \\
\texttt{azimuth}, \texttt{elevation} & Grid (degrees): azimuth $-180$ to 177.5, elevation $-90$ to 90, \SI{2.5}{\degree} steps. \\
\texttt{left/H}, \texttt{right/H}, \texttt{emitter/H} & Complex transfer function $H$ (equation~\ref{eq:ref}), single precision, dimensions [elevation, azimuth, frequency]. HDF5 stores it as a compound of real (\texttt{r}) and imaginary (\texttt{i}) parts. \\
\texttt{<source>/patch\_faces} & Indices (zero-based) of the vibrating faces of \texttt{mesh.stl} (BEM sources only, the piston emitters have attributes radius, centre and axis instead). \\
\texttt{left\_alt1..4}, \texttt{right\_alt1..4}, \texttt{emitter\_alt1..4} & The same for the alternative ear positions and nostril pairs. For the oral emitters, \texttt{emitter\_alt1..5} are the pistons of 2.5 to \SI{6.5}{\milli\meter} radius. \\
attributes & Species and taxon, frame, rotation \texttt{R} ($3 \times 3$ array, $\vec{u}_{\mathrm{CT}} = R\,\vec{u}_{\mathrm{frame}}$), landmarks in the frame (JSON string), number of faces, evaluation radius, conventions, solver version (git commit). The rows at elevation $\pm$\SI{90}{\degree} repeat the single pole value for every azimuth. \\
\bottomrule
\end{tabularx}
\end{table*}

\section*{Data Overview}
\begin{figure*}[!htbp]
\centering
\figfile{0.95\linewidth}{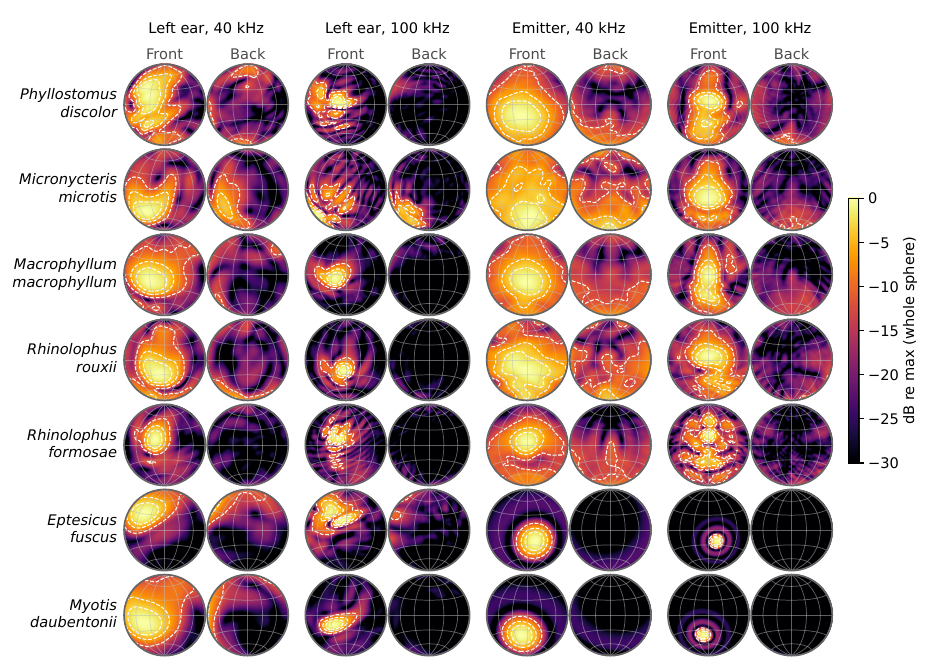}
\caption{Left-Ear and Emitter Patterns of the Seven Species on the Full Sphere. Magnitude at 40 and \SI{100}{\kilo\hertz} in the frame of the dataset (CT frame, for \textit{P. discolor} the pose of De Mey et al.), shown as the front hemisphere (centred straight ahead) and the back hemisphere (centred straight behind, seen from behind the bat), each in Lambert azimuthal equal-area projection. In both discs, the bat's left is on the left and up is up, so the left and right rims are the directions at azimuth \SI{+90}{\degree} and \SI{-90}{\degree}. Graticule every \SI{30}{\degree}. Each pair is in dB relative to its maximum over the whole sphere, with dashed contours at $-3$, $-6$ and \SI{-12}{\decibel}. Nominal source positions. The emitters of the first five species are BEM results for two in-phase nostril patches, those of \textit{E. fuscus} and \textit{M. daubentonii} the analytical pistons calibrated to search calls in free flight in the open field, the narrowest beam of these bats. In the laboratory, in confined spaces and when approaching prey, their beams are considerably wider (see the alternative radii in the dataset and the Usage Notes).}
\label{fig:overview}
\end{figure*}

Fig.~\ref{fig:overview} shows the left-ear and emitter patterns of all species at 40 and \SI{100}{\kilo\hertz} on the full sphere. Supplementary Video 1 (also available at \url{https://youtu.be/Q9Kxj9oIEwg}) shows the seven heads turning with the beam patterns of their left ear and emitter, frequency sweeps of the left ear, the emitter and the two-way ERTF of all species from 20 to \SI{160}{\kilo\hertz}, and the posable models of \textit{M. microtis} and \textit{R. formosae} with their pinnae moving in antiphase. Supplementary Figs.~S5--S11 show both ears, the emitter and the two-way ERTF of every species at six frequencies. Note that the panels are drawn in the frame of each species, whose pitch follows the original scan (for \textit{P. discolor}, the pose of De Mey et al.), so the elevation of the beams should be compared between species with care, and that the emitters of \textit{E. fuscus} and \textit{M. daubentonii} are analytical pistons without a head, whose directivity cannot be compared directly with that of the BEM emitters.

\section*{Technical Validation}
We separate the validation into numerical verification, which asks whether the equations are solved correctly for the given meshes and sources, and comparisons with independent data, which ask how far the modelled responses are from those of real bats. In the first group, we compared the solver with an analytical solution and with NumCalc, studied mesh convergence and angular sampling, and checked the conventions. In the second group, we compared the responses with the earlier simulations of two species and with measurements of \textit{M. daubentonii}. Finally, we quantified the sensitivity to the source assumptions. Table~\ref{tab:validation} summarizes the checks.

\textbf{Metrics:} we compare directivity patterns by their pattern correlation: both magnitudes are converted to dB, normalized to their own maximum over the compared directions and clipped at \SI{-30}{\decibel}, and the Pearson correlation of the two dB patterns is computed with solid-angle weights ($\cos\varphi$ on the grid, uniform on near-uniform direction sets). The rms dB difference uses the same normalized and clipped patterns. The relative complex error $\lVert H - H_{\mathrm{ref}} \rVert / \lVert H_{\mathrm{ref}} \rVert$ over all directions includes level and phase. The directivity index is the peak power over the power averaged over the full sphere, and the half-amplitude angle of a beam is the angular radius of a cone with the same solid angle as the region within \SI{6}{\decibel} of the peak.

\begin{table*}[!tbp]
\centering
\caption{Summary of the Technical Validation. Correlation: pattern correlation of two directivity patterns in dB (median over frequency where a range of frequencies is compared). Relative error: relative complex error over all directions.}
\label{tab:validation}
\small
\renewcommand{\arraystretch}{1.15}
\begin{tabularx}{\linewidth}{>{\raggedright}p{0.2\linewidth} >{\raggedright}X >{\raggedright\arraybackslash}X}
\toprule
Check & Test & Result \\
\midrule
\multicolumn{3}{l}{\textit{Numerical verification}} \\
Analytical solution & Rigid sphere with a vibrating cap, $ka \leq 11$ & \SphereRmsDb rms, \SphereRmsDeg rms in phase \\
Irregular frequency & Sphere, first interior eigenfrequency & CBIE \IrregCbie, Burton-Miller \IrregBm \\
Independent solver & NumCalc, \textit{M. microtis} mesh, 40--\SI{160}{\kilo\hertz} & $\leq$\NumcalcMaxRms rms, $\leq$\SI{1.1}{\degree} rms in phase \\
Discretization & Sphere and subdivided \textit{M. microtis} mesh, 6 to 8 elements per wavelength & Relative error 0.02 (sphere), 0.02--0.06 (head) \\
Geometry resolution & \textit{M. microtis} remeshed at half the edge length & Relative error 0.06--0.22 \\
Mesh convergence & Finer remeshes of three species & Correlation $\geq$0.96 (\textit{M. microtis}), $\geq$0.83 (\textit{P. discolor}, \textit{R. formosae}) \\
Far field & Response at \SI{1}{\meter} vs.\ \SI{20}{\meter} & Correlation $\geq$\FarCorr \\
Angular sampling & Interpolation of the \SI{2.5}{\degree} grid & $\leq$\InterpRmsMax rms \\
Conventions & Group delays vs.\ landmark geometry & Correlation \EmDelayCorr (emitter), \ItdCorr (interaural) \\
\midrule
\multicolumn{3}{l}{\textit{Comparison with independent data}} \\
Earlier simulations & \textit{M. microtis} and \textit{P. discolor}, after landmark calibration & Correlation \RefMmicrL--\RefMmicrE (\textit{M. microtis}), \RefPdiscL--\RefPdiscE (\textit{P. discolor}) \\
Measured HRTF & \textit{M. daubentonii}, another specimen & Correlation \VhCorrOurs (Mesh2HRTF: \VhCorrTheirs), 0.82 (0.81) after alignment \\
Placement rule & \textit{M. daubentonii}, measured HRTF & 98th--99th percentile of 211 candidates \\
In-flight beam & \textit{M. daubentonii} & Piston \HbHalfDiff narrower \\
\midrule
\multicolumn{3}{l}{\textit{Sensitivity}} \\
Source placement & Alternative ear positions and nostril pairs & Correlation \EnsEarCorr (ears), \EnsEmCorr (nostrils) \\
Piston radius & Radius $\pm$\SI{20}{\percent} & Directivity index $\pm$\PistonDiTwenty dB \\
Pinna posture & \textit{M. microtis}, flex 10 to \SI{25}{\degree} & Correlation 0.73--0.92 \\
\bottomrule
\end{tabularx}
\end{table*}

\subsection*{Solver verification}
\begin{figure*}[!htbp]
\centering
\figfile{\linewidth}{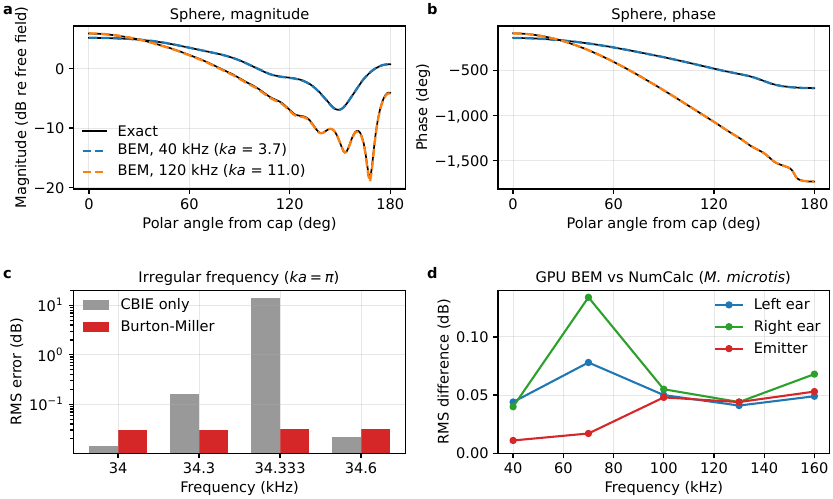}
\caption{Verification of the Solver. Panels a and b show the magnitude and phase of the field of a vibrating cap on a rigid sphere (radius \SI{5}{\milli\meter}) at 40 and \SI{120}{\kilo\hertz} against the polar angle, for the exact series solution (solid black) and the BEM (dashed). Panel c shows the rms error at and around the first irregular frequency of a 5,120-element sphere, for the conventional boundary integral equation (CBIE) alone and for the Burton-Miller formulation (logarithmic scale). Panel d shows the rms difference between our GPU BEM and NumCalc (Mesh2HRTF) for the two ears and the emitter of the \textit{M. microtis} dataset mesh. The BEM matches the analytical solution and NumCalc closely, and the Burton-Miller formulation removes the failure of the CBIE at the irregular frequency.}
\label{fig:validation}
\end{figure*}

A rigid sphere of radius \SI{5}{\milli\meter} (20,480 elements) with a vibrating spherical cap (radius \SI{0.8}{\milli\meter}, twice the ear patch) has an exact series solution for the radiated field. The BEM agrees with it to within \SphereRmsDb rms and \SphereMaxDb maximum in magnitude, and \SphereRmsDeg rms in phase, up to $ka = 11$ (\SI{120}{\kilo\hertz}, Fig.~\ref{fig:validation}a,b). This test also verifies the referencing of equation~\ref{eq:ref} and the time convention, as the series solution is computed independently of the BEM. At the first irregular frequency of the 5,120-element sphere (\SI{34.333}{\kilo\hertz}, $ka \approx \pi$), the CBIE alone fails with an rms error of \IrregCbie (\SI{21}{\decibel} maximum), while the Burton-Miller formulation stays at \IrregBm (Fig.~\ref{fig:validation}c). As the irregular frequencies of a bat-sized head are dense in the ultrasonic band, the conventional formulation of the BEM of Otani and Ise \citep{otaniFastCalculationSystem2006}, used for the earlier bat simulations \citep{demeySimulatedHeadRelated2008, vanderelstWhatNoseleavesFM2010}, is prone to such failures throughout the band, whereas the Burton-Miller formulation of our solver avoids them.

NumCalc of Mesh2HRTF is an established open-source BEM for HRTFs \citep{ziegelwangerMesh2HRTFOpensourceSoftware2015, kreuzerFastMultipoleBoundary2009}, whose results have been validated against acoustic measurements on humans \citep{ziegelwangerNumericalCalculationListenerspecific2015}. We computed the nominal ear and emitter sources of the \textit{M. microtis} dataset mesh with NumCalc (Mesh2HRTF 1.3.0) at five frequencies from 40 to \SI{160}{\kilo\hertz}. The two solvers agree to within \NumcalcMaxRms rms in magnitude and \SI{1.1}{\degree} rms in phase, with pattern correlations of 0.9999 or higher (Fig.~\ref{fig:validation}d). This comparison was made on the frontal hemisphere. On the full sphere, at 80, 120 and \SI{160}{\kilo\hertz}, the relative complex difference between the two solvers on this mesh is at most 0.005. As both use the same discretization, this shows that our implementation solves the same discrete problem as NumCalc, but it does not test the discretization error, which the next section addresses.

\subsection*{Discretization, far field and angular sampling}
Supplementary Note 5, Supplementary Table~S2 and Supplementary Fig.~S2 describe the mesh convergence, far-field and angular sampling checks in detail, and we summarize them here. At six elements per wavelength, the density of the production meshes at the top of their bands, the BEM reproduces the analytical sphere with a relative complex error of about 0.02 (0.2 to \SI{0.3}{\decibel} rms). On the \textit{M. microtis} mesh, the discretization error alone, measured against NumCalc on the same mesh with every triangle subdivided into four, is 0.04 to 0.06 for the ear and 0.02 to 0.03 for the emitter, but a new remesh of the CT surface at half the edge length differs by 0.12 to 0.22 and 0.06 to 0.14. At high frequencies, the uncertainty of the data is therefore dominated by how faithfully the mesh represents the fine anatomy, rather than by the number of elements per wavelength. For \textit{P. discolor} and \textit{R. formosae}, whose meshes are at the face budget and whose source patches contain only 4 to 11 elements, different remeshes differ more, with pattern correlations down to 0.83 above \SI{120}{\kilo\hertz}, so the parts of their bands above about \SI{100}{\kilo\hertz} carry a larger uncertainty. The magnitudes at \SI{1}{\meter} represent the far-field directivity (pattern correlations of at least \FarCorr with the response at \SI{20}{\meter}), while the phase still contains the curvature of the wavefront over the head (relative complex differences of up to \FarErr, corresponding to delays well below a microsecond). Finally, bilinear interpolation of the \SI{2.5}{\degree} grid reproduces directly computed magnitudes to within \InterpRmsMax rms above \SI{-30}{\decibel}.

\subsection*{Conventions and registration}
A sign error in the azimuth, a mirrored mesh, a misplaced landmark or a wrong time convention would leave the magnitudes plausible but make the data wrong. To catch such errors, we compared the group delays of the data, fitted from the phase slope between 20 and \SI{40}{\kilo\hertz}, with the geometric delays of the landmarks (Supplementary Note 6 and Supplementary Table~S3). The emitter delays correlate with those of the nostril midpoint or the mouth with \EmDelayCorr (\textit{R. formosae}: \EmDelayCorrRform, as its horseshoe moves the apparent acoustic centre of the emission forward of the nostrils, onto the midline), and the interaural group-delay differences correlate positively with the geometric ones for all species (\ItdCorr, Fig.~\ref{fig:sensitivity}c), as required for a correct azimuth and time convention.

\subsection*{Comparison with earlier simulations}
\begin{figure*}[!htbp]
\centering
\figfile{\linewidth}{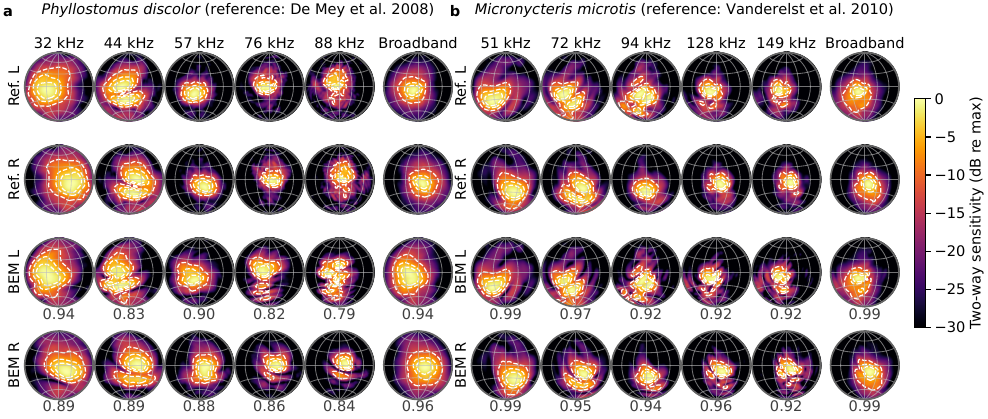}
\caption{Comparison with Earlier Simulations. Two-way ERTF (emitter magnitude times ear magnitude) of \textit{P. discolor} (left block, reference: De Mey et al. \citep{demeySimulatedHeadRelated2008}) and \textit{M. microtis} (right block, reference: Vanderelst et al. \citep{vanderelstWhatNoseleavesFM2010}) in Lambert azimuthal equal-area projection of the frontal hemisphere (centre straight ahead, the bat's left on the left, graticule every \SI{30}{\degree}). Rows: reference (Ref.) and this dataset (BEM), left (L) and right (R) ear. Columns: five frequencies and the broadband pattern (call-weighted for \textit{P. discolor}). Each map is in dB relative to its maximum, with contours at $-3$, $-6$ and \SI{-12}{\decibel}. The number under each map of this dataset is its pattern correlation with the reference.}
\label{fig:reference}
\end{figure*}

The ERTFs of \textit{P. discolor} \citep{demeySimulatedHeadRelated2008} and \textit{M. microtis} \citep{vanderelstWhatNoseleavesFM2010} were computed earlier from the same CT scans, with the BEM of Otani and Ise \citep{otaniFastCalculationSystem2006, otaniFastCalculationMethod2003}, a conventional boundary integral equation without measures against irregular frequencies, on meshes of fewer than 32,000 triangles. As our landmarks for these two species were chosen to match these simulations best (see Methods), the agreement is a consistency check with an independent solver and meshing, not a validation. For \textit{M. microtis}, the two-way ERTF agrees with pattern correlations of 0.92 to 0.99 between 51 and \SI{149}{\kilo\hertz}, and the median pattern correlations over the reference band are \RefMmicrL, \RefMmicrR and \RefMmicrE for the left ear, the right ear and the emitter (Fig.~\ref{fig:reference}, Supplementary Fig.~S4). For \textit{P. discolor}, they are \RefPdiscL, \RefPdiscR and \RefPdiscE, limited by the reference itself, which is not smooth in frequency (patterns \SI{0.5}{\kilo\hertz} apart correlate with a median of only \RefPdiscSmooth, against \RefMmicrSmooth for the \textit{M. microtis} reference), as expected from irregular-frequency failures and a coarse mesh. De Mey et al. compared their simulation with the HRTFs measured by Firzlaff and Schuller on other specimens of \textit{P. discolor} \citep{firzlaff2003} and found the region of highest sensitivity and the spectral and interaural patterns in good agreement, within the variability between individuals \citep{demeySimulatedHeadRelated2008}. As our \textit{P. discolor} transfer functions agree with that simulation (Fig.~\ref{fig:reference}), this agreement extends to them, although we did not compare them with the measurements directly.

\subsection*{Comparison with measurements}
\begin{figure*}[!htbp]
\centering
\figfile{\linewidth}{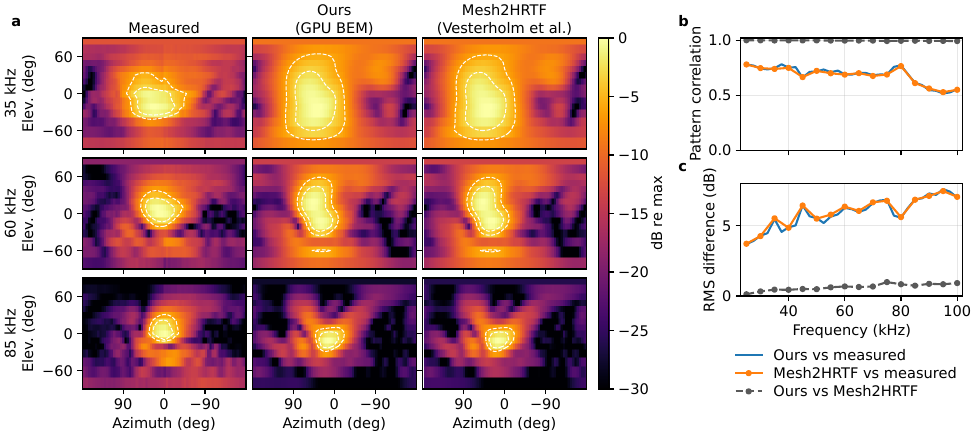}
\caption{Comparison with a Measured HRTF of \textit{M. daubentonii}. Left-ear HRTF of specimen Mdau2 of Vesterholm et al.~\citep{vesterholm2025}. Panel a shows maps of the measured in-ear HRTF, of their Mesh2HRTF simulation and of our simulation on the same micro-CT scan. Panels b and c show the pattern correlation and the rms dB difference between them against frequency.}
\label{fig:measurement}
\end{figure*}

The specimens of this dataset can no longer be measured. To compare the modelling pipeline with measurements, we used the public data of Vesterholm et al. \citep{vesterholm2025, vesterholm2025data}, who measured the in-ear HRTF of a dead \textit{M. daubentonii} between 25 and \SI{100}{\kilo\hertz}, scanned the same specimen with micro-CT and simulated it with Mesh2HRTF. We meshed their scan with our pipeline and computed the HRTF of the same ear with our solver at their measurement directions, comparing magnitudes only, as their measured impulse responses are minimum-phase. Our simulation reproduces their independent Mesh2HRTF simulation closely (median pattern correlation \VhCorrSim), so our solver and meshing perform as the established Mesh2HRTF does. Both simulations, however, deviate from the measurement in the same way (median pattern correlations of \VhCorrOurs and \VhCorrTheirs over the 16 frequencies at which both are available, Fig.~\ref{fig:measurement}a--c), with main lobes about \SI{25}{\degree} apart. This shared difference can come from the alignment of the measurement with the mesh, from the mesh itself, or from what both simulations share, such as the rigid surface and the receiver at the ear canal entrance. To test the alignment, we fitted an additional rotation between the measurement directions and the mesh. Both simulations independently require the same rotation of about \SI{26}{\degree}. With it, their median pattern correlations with the measurement increase from \VhCorrOurs and \VhCorrTheirs to 0.82 and 0.81, and the main lobes lie \SI{11}{\degree} apart. A large part of the difference is therefore consistent with an offset between the measured directions and the scan, from the registration or from a change in the posture of the dead specimen between measurement and scan, rather than with an error of the simulation. The remainder cannot be attributed further with these data. This comparison thus shows that our pipeline reproduces a measured HRTF as well as Mesh2HRTF does. It does not validate either simulation independently against ground truth.

The same data allow an independent test of the receiver placement rule, which was derived from \textit{P. discolor} and \textit{M. microtis} only. On the mesh of Vesterholm et al., the rule chooses, among 211 candidate positions around the ear canal entrance, a position \SI{1.0}{\milli\meter} from their receiver. Its pattern correlation with the measured HRTF (0.69 with their alignment, 0.78 with the fitted one) lies at the 98th to 99th percentile of all candidates, within 0.01 of the best candidate and of their own receiver position, whereas the median candidate reaches only 0.13 to 0.38. The most occluded candidate alone performs similarly here (0.68 and 0.77), so this test confirms the occlusion component of the rule in particular.

For the emitter, we compared the piston of \textit{M. daubentonii} with the in-flight emission beam of that species measured by H\"afele et al. \citep{hafele2026, hafele2026data}, which is independent of the field data used for calibration. The in-flight beam has the shape of a piston beam but is wider: between 45 and \SI{75}{\kilo\hertz}, our piston is on average \HbHalfDiff narrower, and a radius of about \HbRadius fits the in-flight beam, close to the \SI{3.1}{\milli\meter} that corresponds to the laboratory beam of this species \citep{surlykkeEcholocatingBatsEmit2009}. Our calibrated pistons therefore represent the narrow open-field search beam, and bats in confined spaces are better described by the alternative radii of about \SI{3.5}{\milli\meter}. The laboratory beam of \textit{E. fuscus} is wider than its field beam as well \citep{hartley1989}. Supplementary Note 7 and Supplementary Table~S3 give the details of both comparisons.

\subsection*{Sensitivity to the source assumptions}
\begin{figure*}[!htbp]
\centering
\figfile{\linewidth}{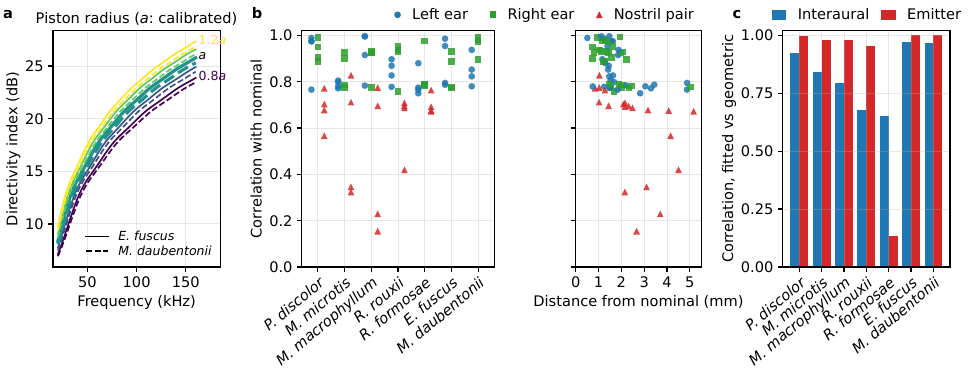}
\caption{Sensitivity to the Source Assumptions and Check of the Conventions. Panel a shows the directivity index of the piston emitters of \textit{E. fuscus} and \textit{M. daubentonii} for 0.8 to 1.2 times the calibrated radius $a$ (thick: calibrated). Panel b shows the pattern correlation of every alternative left-ear position (circles), right-ear position (squares) and nostril pair (triangles) with the nominal one, median over frequency, per species (left) and against the distance from the nominal position (right, nostril pairs: distance between the pair midpoints). Panel c shows the correlation between fitted and geometric group delays (20--\SI{40}{\kilo\hertz}): the interaural difference of the ears (blue, full sphere) and the emitter delay (red, within \SI{10}{\decibel} of the maximum, for the two piston emitters it holds by construction).}
\label{fig:sensitivity}
\end{figure*}

\begin{figure*}[!htbp]
\centering
\figfile{\linewidth}{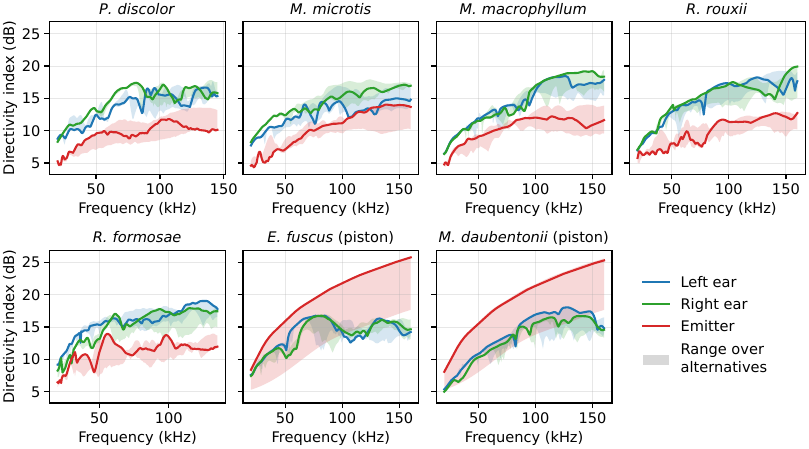}
\caption{Directivity Index against Frequency. Directivity index (peak power over the power averaged over the full sphere) of the left ear (blue), the right ear (green) and the emitter (red) of each species, for the nominal source positions (lines) and the range over the alternative positions (shaded). For the two vespertilionids, the shaded range spans the pistons of 2.5 to \SI{6.5}{\milli\meter} radius.}
\label{fig:directivity}
\end{figure*}

The placement of the sources is the largest modelling choice in the dataset. Over frequency, the alternative ear positions give patterns that correlate with the nominal one with medians of \EnsEarCorr, and the alternative nostril pairs with \EnsEmCorr, the lowest values belonging to nostril pairs 2 to \SI{4.5}{\milli\meter} from the nominal pair (Fig.~\ref{fig:sensitivity}b). The emitters are thus more sensitive to the placement than the ears, and Fig.~\ref{fig:directivity} shows the resulting range of the directivity index. For the piston emitters, a radius \SI{20}{\percent} smaller or larger changes the directivity index by a median of \PistonDiTwenty~dB (Fig.~\ref{fig:sensitivity}a). Users who need another gape or orientation can evaluate equation~\ref{eq:piston} with the released code.

\subsection*{Posed models}
\begin{figure*}[!htbp]
\centering
\figfile{\linewidth}{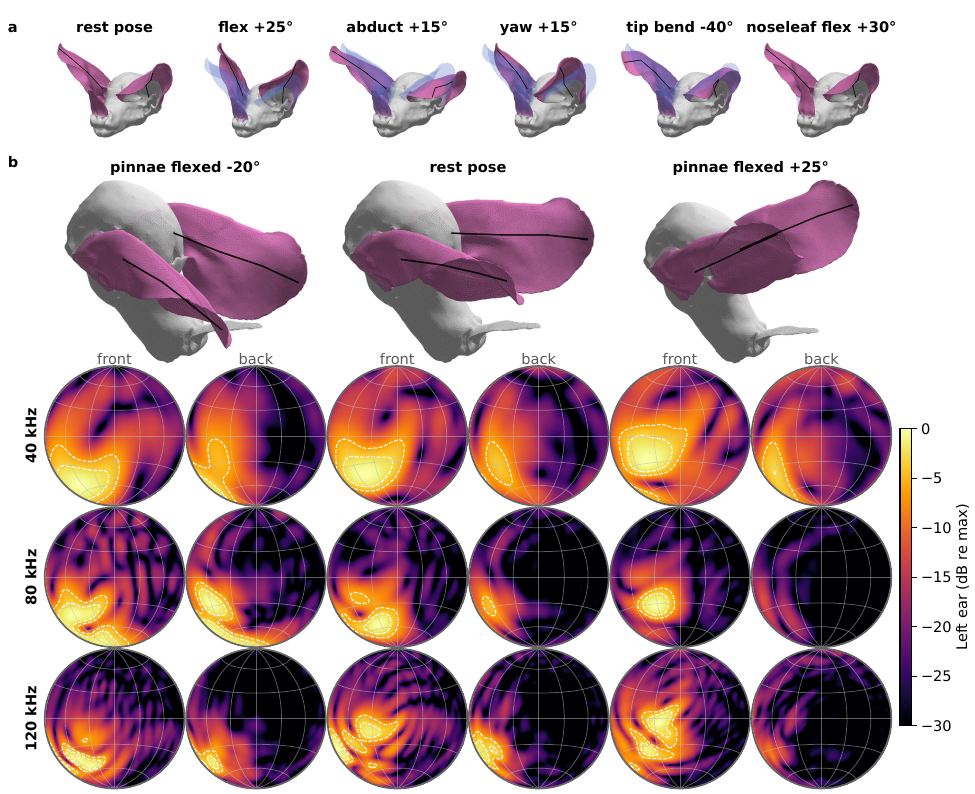}
\caption{Posable Head Model of \textit{M. microtis}. Panel a shows the posing options on the rigged head: the rest pose, and both pinnae flexed backward, abducted, turned forward (yaw), bent at the tip, and the noseleaf flexed, all seen from the same direction with the head levelled for display (the line from the ears to the nostrils horizontal), with the rigged parts in magenta, their rest position in light blue, and the bone chains of the pinnae in black. Panel b shows, for both pinnae flexed \SI{20}{\degree} forward, the rest pose and both pinnae flexed \SI{25}{\degree} backward, the posed head above the left-ear transfer function at 40, 80 and \SI{120}{\kilo\hertz}, as front and back hemispheres in Lambert azimuthal equal-area projection (dB re maximum, contours at $-3$ and \SI{-6}{\decibel}). Every pose was computed with the BEM, and nothing is interpolated.}
\label{fig:posing}
\end{figure*}

To show what the posable models offer, and to check that posed meshes remain valid inputs for the BEM, we flexed both pinnae of \textit{M. microtis} about their base joints from $-20$ to \SI{+25}{\degree} in \SI{1.5}{\degree} steps and computed both ears and the emitter at 40, 80 and \SI{120}{\kilo\hertz} for every pose (Fig.~\ref{fig:posing}b, Fig.~\ref{fig:posing}a shows the other posing options). All 31 posed meshes passed the validity check, and the ear landmarks moved by less than \SI{2}{\micro\meter}, so that the source patches kept their triangles. Poses within \SI{1}{\degree} of the rest pose reproduce the dataset with pattern correlations of at least 0.995. The ear patterns change markedly with the posture of the pinnae: at 10 to \SI{25}{\degree} of flex, the left-ear pattern correlates with that of the rest pose with only 0.73 to 0.92, its main lobe moves by up to \SI{18}{\degree} at \SI{40}{\kilo\hertz}, and at 80 and \SI{120}{\kilo\hertz} the strongest lobe changes. Within the valid ranges of the joints, which allow flex and abduction of the pinna base by about $\pm$15 to \SI{\pm 40}{\degree} and yaw by $\pm$15 to \SI{\pm 35}{\degree}, depending on the species (Supplementary Note 8 and Supplementary Table~S4), posture thus changes the directivity at least as much as the source placement does.

\section*{Usage Notes}
The HDF5 files can be read with any HDF5 library. The complex values are stored as a compound of real and imaginary parts, which h5py returns as complex numbers and MATLAB's \texttt{h5read} as a structure with fields \texttt{r} and \texttt{i}. The SOFA files can be read with the SOFA toolboxes for MATLAB/Octave and Python \citep{majdakSpatiallyOrientedFormat2013}. The released code \citep{thiscode} contains a small Python module that loads a species folder, interpolates the transfer functions to arbitrary directions, rotates the head, computes impulse responses and simulates the binaural echoes of point reflectors, and the dataset contains examples of these operations in Python and MATLAB.

A few practical notes apply. (1) The phase of $H$ changes quickly with direction at high frequencies, as it contains the delay of the source relative to the origin. For interpolation between grid directions, multiply $H$ by $e^{-jk\,\vec{u}\cdot\vec{p}}$ with $\vec{p}$ the source landmark, interpolate the real and imaginary parts, and restore the delay (as the released code does), or interpolate magnitude only. (2) Impulse responses follow from an inverse real FFT over frequency, after the band below \SI{20}{\kilo\hertz} and above the top of the band has been filled in, for instance with a smooth taper to zero (the released code uses a \SI{5}{\kilo\hertz} raised-cosine taper at both band edges). They are referenced to the origin and need a small pre-delay (the released code uses \SI{0.2}{\milli\second}), as they start up to \SI{60}{\micro\second} before the arrival at the origin. (3) The two-way ERTF of a direction is the product of the emitter and the ear transfer functions. The echo of a point reflector at distance $r$ additionally contains the spreading loss, the round-trip delay $2r/c$ and the atmospheric absorption, which the data do not include. (4) The data are responses at \SI{1}{\meter}: their magnitudes represent the far-field directivity, while their phase still contains the curvature of the wavefront over the head at that distance (see Technical Validation). For reflectors closer than about \SI{1}{\meter}, the incident field is not plane over the head and the data are an approximation. (5) For \textit{P. discolor} and \textit{R. formosae}, whose source patches are resolved by only 4 to 11 elements, the responses above about \SI{100}{\kilo\hertz} depend noticeably on the mesh realization. (6) The source positions are modelling choices, and the transfer functions are very sensitive to them: alternative positions 0.5 to \SI{5}{\milli\meter} from the nominal ones correlate with them with \EnsEarCorr for the ears and \EnsEmCorr for the nostril pairs (Fig.~\ref{fig:sensitivity}b). The alternative positions show how much a result depends on this choice, which is most at high frequencies and for the emitters. Results that depend on fine spectral or spatial details should therefore be checked against the alternatives. (7) The nominal oral emitters of \textit{E. fuscus} and \textit{M. daubentonii} are calibrated to search calls in free flight in the open field, the narrowest beam these bats emit. They are therefore not a general emission pattern of these species, and users should adapt the width of the beam to the situation they model. In the laboratory, in confined spaces and during the approach to prey, the beam is considerably wider, which corresponds to piston radii of about 3 to \SI{5}{\milli\meter} (Methods). The dataset provides the pistons of 2.5 to \SI{6.5}{\milli\meter} radius (\texttt{emitter\_alt1..5}), and the released code computes the piston for any radius (\texttt{ERTF.piston}).

\textbf{Posing the models:} a pose is a set of joint rotations in degrees. The released code \citep{thiscode} applies it to the rest mesh, checks the result, moves the landmarks with the surface and builds the source patches, after which the posed mesh can be solved like the dataset meshes, for example:
{\scriptsize
\begin{verbatim}
from batertf import rig, sources, project, gpubem, grid
r = rig.load_rig("mmicr")             # skeleton and weights
pose = {"pinna_both.base": {"flex": 15,   # pinnae backward
                            "yaw": 10},   # and forward
        "noseleaf.base": {"flex": -10}}   # noseleaf forward
m = r.deform(pose)                    # posed mesh
assert r.check(m, pose)["valid"]      # valid surface
lm = r.deform_landmarks(pose, m)      # landmarks on it
patch = sources.disc_patch(m, lm["ear_left"],
                           project.EAR_RADIUS)
pts = grid.unique_points("sphere")[0]
H = gpubem.solve_lu(m, {"left": [patch]}, [80e3], pts)
\end{verbatim}
}
Poses can also be set interactively: the glTF model in \texttt{rig/} opens in 3D software such as Blender, and a posed model exported from there is converted back into joint rotations with \texttt{rig.\allowbreak pose\_from\_gltf}. Each posed frequency costs one LU decomposition (about \SI{20}{\second} for 50,000 elements on our GPU). A posed mesh can also be computed with Mesh2HRTF on a CPU (the released code writes the NumCalc project with the patch faces, \texttt{batertf/\allowbreak numcalc.py}). The released code includes a step-by-step guide (\texttt{examples/POSING.md}), an example that poses a head, computes it and compares it with the rest pose, and a tool that computes grids of poses on one or more GPUs. We recommend staying within the valid joint ranges of Supplementary Note 8, checking every combined pose, and keeping in mind that the skinning is a geometric deformation, not a model of the tissue.

\textbf{Further specimens:} the released pipeline (mesh cleaning, source placement, solver and rig) can be applied to other head meshes. For example, the micro-CT and photogrammetry meshes of \textit{M. daubentonii} and \textit{M. nattereri} of Vesterholm et al. \citep{vesterholm2025, vesterholm2025data}, released under CC BY 4.0, can be processed in the same way, as we did for one of them in the Technical Validation.

The limitations of the dataset follow from its construction. The heads are rigid: the impedance of the tissue and fur, the ear canal and middle ear, and the vocal tract are not modelled. Each species is represented by one preserved specimen, whose pinnae and noseleaf may be deformed by preservation and whose ear posture is fixed, whereas live bats move their pinnae and, in rhinolophids, their noseleaf \citep{vanderelst2011}. The oral emitters are the most approximate part of the dataset. Real bats emit through the open mouth, whereas the meshes have closed mouths, so these emitters are analytical pistons that approximate the open mouth, not BEM models of it, and they omit the head. Their beam is symmetric about the piston axis, whereas the laboratory beam of \textit{E. fuscus} shows a ventral secondary lobe \citep{hartley1989}, and the beam of a real bat varies with its behaviour. The five alternative radii span the range of published beam widths, but besides the nominal radius, only the \SI{3.5}{\milli\meter} piston of \textit{M. daubentonii} matches a measured beam of a specific behaviour (in flight in a wind tunnel). The placement rule for the nostrils failed for one of the five nasal species (\textit{R. formosae}), where we used its second choice instead. Finally, the bands of \textit{P. discolor} and \textit{R. formosae} end at 145 and \SI{135}{\kilo\hertz}.

\section*{Data Availability}
The dataset is available at Zenodo \citep{thisdataset} (\url{https://doi.org/10.5281/zenodo.23729269}) under CC BY 4.0: per species, the clean head mesh, the transfer functions of both ears and of the emitter (nominal and alternative sources) as HDF5 and SOFA files, the landmarks, and the metadata, as described in the Data Records. The earlier reference simulations used for comparison \citep{demeySimulatedHeadRelated2008, vanderelstWhatNoseleavesFM2010} are not part of this dataset. The measurements used in the Technical Validation are available from their own repositories \citep{vesterholm2025data, hafele2026data}, under CC BY 4.0 and CC BY-NC-ND 4.0, respectively, and neither is redistributed with this dataset.

\section*{Code Availability}
The code that produced the dataset (mesh cleaning, GPU BEM, source placement, production, validation and figures) and the Python and MATLAB examples are available on GitHub (\url{https://github.com/Cosys-Lab/bat_ertf_dataset}) and archived together with the dataset at Zenodo \citep{thisdataset} (version 1.0.0, \texttt{bat\_ertf\_code-1.0.0.zip}), under the BSD 3-Clause licence. Loading and using the data (\texttt{batertf/dataset.py} and the Python examples) requires only Python 3 with NumPy and h5py (Matplotlib for the plots), and the MATLAB examples base MATLAB (R2017b or newer, with the Signal Processing Toolbox for the spectrograms of the echo example). Recomputing the data requires Python 3.12, PyTorch 2.14 with CUDA, NumPy, SciPy, h5py, trimesh, PyMeshLab 2025.7 and sofar 1.3, and an NVIDIA GPU with \SI{48}{\giga\byte} of memory for meshes of 58,000 elements (smaller GPUs suffice for smaller meshes). The comparison with NumCalc used Mesh2HRTF 1.3.0 \citep{ziegelwangerMesh2HRTFOpensourceSoftware2015, brinkmann2023}.

\bibliographystyle{unsrtnat}
\bibliography{refs}

\section*{Author Contributions}
J.S. conceived the study, prepared the meshes, developed the solver, the source placement and the posable models, computed and validated the dataset, and wrote the manuscript.

\section*{Competing Interests}
The author declares no competing interests.

\section*{Acknowledgements}
The original head meshes were created as part of the EU projects Cilia (FP6-IST-016039) and ChiRoPing (FP7-215370).

\section*{Funding}
The author received no specific funding for this work.

\onecolumn
\includepdf[pages=-]{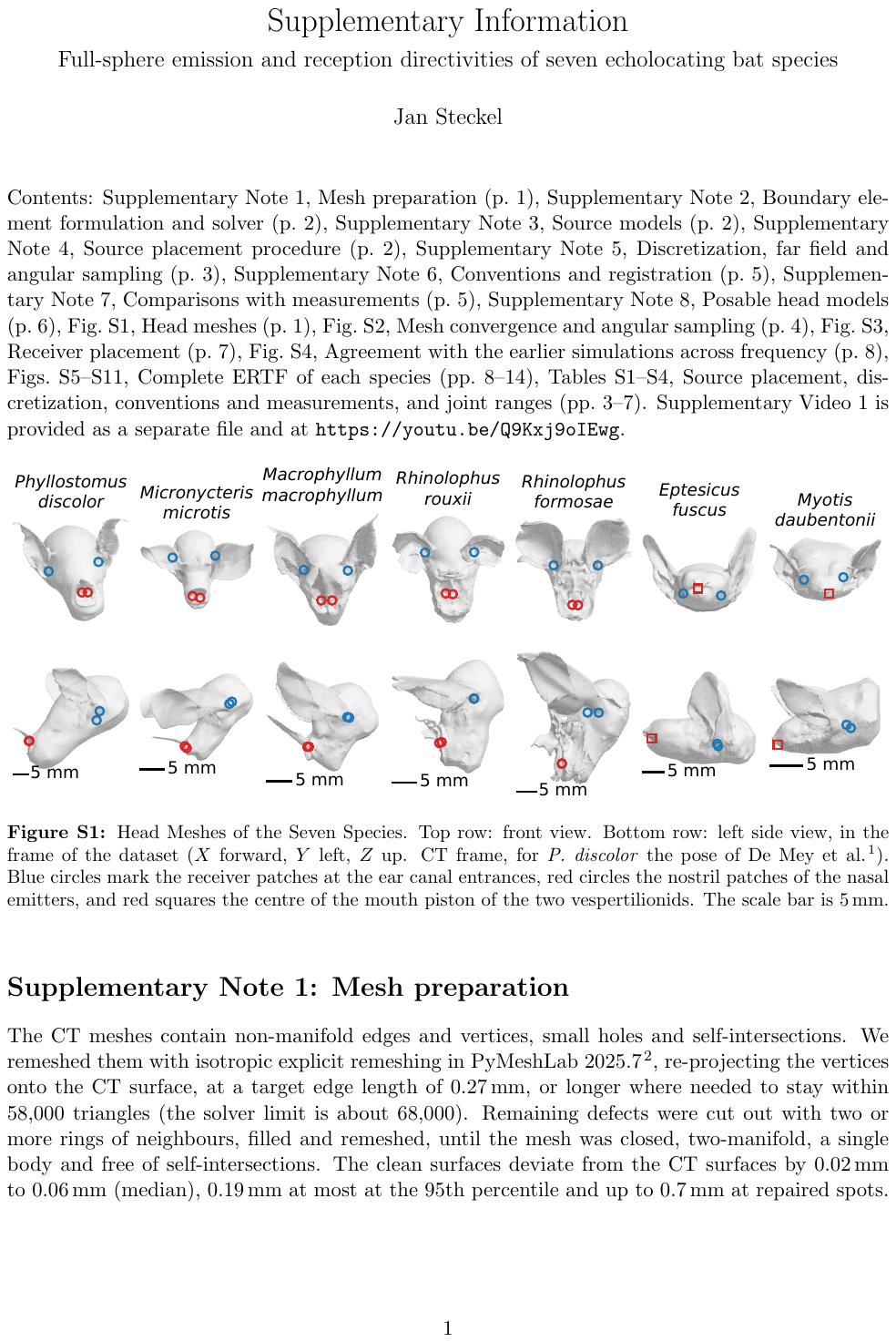}
\end{document}